\documentstyle[11pt,amsmath,amssymb,epsf,epic,cite]{article}
\numberwithin{equation}{section}

\newcommand{\be}{\begin{equation}}
\newcommand{\ee}{\end{equation}}
\newcommand{\bea}{\begin{eqnarray}}
\newcommand{\eea}{\end{eqnarray}}

\renewcommand{\c}{{\rm c}}
\newcommand{\e}{{\rm e}}

\renewcommand{\d}{{\rm d}}

\newcommand{\grintl}{[\kern-.18em [}
\newcommand{\grintr}{]\kern-.18em ]}

\newcounter{resultcounter}[section]

\newtheorem{prop}[resultcounter]{Proposition}

\newtheorem{definition}[resultcounter]{Definition}

\def\bed{\begin{definition}}
\def\eed{\end{definition}}

\newcommand{\r}{{\rm R}}

\newcommand{\rx}{{\mathbb R}}
\renewcommand{\i}{{\rm i}}

\newcommand{\fer}[1]{(\ref{#1})}
\newcommand{\scalprod}[2]{\left\langle {#1}, {#2}\right\rangle}

\begin{document}

\title{Creation of Two-Particle Entanglement in Open Macroscopic Quantum Systems\\}

\author{M. Merkli\,\footnote{Department of Mathematics and Statistics,  Memorial
University of Newfoundland, St. John's, Newfoundland, Canada A1C
5S7.}\ \footnote{Supported by NSERC under Discovery Grant 205247.
Email: merkli@mun.ca, http://www.math.mun.ca/$\sim$merkli/} \and  G.P.
Berman,\footnote{Theoretical Division, MS B213, Los Alamos National
Laboratory, Los Alamos, NM 87545, USA. Email: gpb@lanl.gov. Work was
carried out under the auspices of  the NNSA of the U. S. DOE at LANL
under Contract No. DEAC52-06NA25396.  Supported by the LDRD program at LANL.}
\and F. Borgonovi,\footnote{Dipartimento di Matematica e
Fisica, Universit\`a Cattolica, via Musei 41, 25121 Brescia, Italy.
Email: fausto.borgonovi@unicatt.it, http://www.dmf.unicatt.it/$\sim$borgonov}
\footnote{INFN, Sezione di Pavia, Italy.}\and and V.I. Tsifrinovich\footnote{Department of Applied Physics, Polytechnic Institute of NYU, 6 MetroTech Center, Brooklyn, NY 11201, USA. Email: vtsifrin@poly.edu}}

\date{\today}
\maketitle
\vspace{-.5cm}

\markright{righthead}{LA-UR 11-05003}

\begin{abstract}
We consider an open quantum system of $N$ not directly interacting spins (qubits) in contact with both local and collective thermal environments.
The qubit-environment interactions are energy conserving. We trace out the
 variables of the thermal environments and $N-2$ qubits to obtain the time-dependent reduced density matrix for two arbitrary qubits. We numerically simulate the reduced dynamics and the creation of entanglement (concurrence) as a function of the parameters of the thermal environments and the number of qubits, $N$. Our results demonstrate that the two-qubit entanglement generally decreases as $N$ increases. We show analytically that in the limit $N\rightarrow\infty$, no entanglement can be created. This indicates that collective thermal environments cannot create two-qubit entanglement when many qubits are located within a region of the size of the environment coherence length. We discuss possible applications of our approach to the development of a new quantum characterization of noisy environments.
\end{abstract}

\thispagestyle{empty}
\setcounter{page}{1}
\setcounter{section}{1}
\setcounter{section}{0}

\section{Introduction}
In  open many-body systems, such as solid-state and biological ones, macroscopic quantum behavior reveals itself in many ways. Often the quantitative parameter used to measure ``quantumness'' (possibly of macroscopic order) is entanglement. The presence of entanglement implies that the wave function (or the reduced density matrix) cannot be represented as a product  of the corresponding objects for the individual qubits. It is important to note that to produce and to measure entanglement in such systems, one does not necessarily need to know much detail about the system, possibly not even its Hamiltonian \cite{witness}. The questions then are: How useful is entanglement as a measure of quatumness,  what can it add to our knowledge of system properties and behavior, and how can it be utilized? Indeed, just knowing that the system is entangled (knowledge of complicated quantum behavior) is not sufficient to imply that its quantum properties are useful for specific applications. Fortunately, however, in some situations entanglement provides very useful properties, including  additional exponential resources for quantum computation \cite{chuang} and a possible enhancement of  photosynthesis in bio-systems \cite{photo}. (See also references therein.)

Entanglement could be produced by direct interaction between the qubits. This interaction should be of a ``conditional'' nature, mixing an initial product state (disentangled) in such a way that the final state becomes correlated in a quantum way. Many aspects of entanglement creation are widely discussed in the literature. (See, for example, \cite{Braun,YE2,BLC,HZ,PR,MBBG} and references therein.)

More recently, interest appeared in the possibility to create entanglement in the absence of direct interactions between qubits (or when the latter are very small). Entanglement can then still be created merely by the indirect interaction of non-interacting qubits through a collective thermal bath. In \cite{Braun} this situation was considered for a model of two non-interacting spins $1/2$ (qubits) interacting only with a collective thermal bosonic environment. It was demonstrated numerically in \cite{Braun} that for some initially unentangled two-qubit states, and under some conditions on the the thermal bath, measurable entanglement between the two qubits is created for intermediate times. The model of \cite{Braun} is energy-conserving, ignoring relaxation processes for the qubits, and including only the effects of decoherence. In \cite{MBBG} these results were extended to a more general model having (i) both local and collective thermal environments (at the same temperature) and (ii) energy conserving and energy exchange interactions between qubits and their environments. The conditions for entanglement creation were discussed and analyzed numerically in \cite{MBBG}. It was concluded that, in spite of the competition between the local thermal environments (which destroy entanglement) and the collective thermal environment (tending to create entanglement), the creation of measurable entanglement can be realized for some finite times. In both papers, \cite{Braun} and \cite{MBBG}, only two qubits are analyzed, hence no direct connection to many-body systems was made. As was recently shown in \cite{MBBR}, the presence of a large number of indirectly interacting qubits interacting only with their common collective thermal environment could significantly modify the effective single-qubit characteristics including their relaxation and decoherence rates.

\medskip

In the present paper we propose a new approach to the analysis of thermal environments, namely, the analysis of their ability to create quantum entanglement. For example, in biological systems, noise produced by a thermal environment can usually be described using standard noise characteristics, including space and time correlation functions (with the corresponding correlation lengths), the spectral density of noise, and high order correlations (if needed). These characteristics allow one to introduce, estimate, and measure important parameters of the biological system, such as relaxation and decoherence (dephasing) times, kinetic and transport coefficients and others. We propose to take the next step in this direction: to develop a new method for characterizing noisy environments created by biological species at room temperature -- a quantum
characterization of noisy environment.

The basic parameter measured in our method is the two-qubit (two-spin) concurrence describing the level of two-qubit entanglement created by noisy environments at room temperature. Experimentally, the method can be implemented using the well-developed technique of liquid state nuclear magnetic resonance (NMR) quantum computers, using an ensemble of two-qubit systems. The qubits are represented by nuclear spins in molecules which are diluted in a liquid. The method we propose will allow one to determine the level of entanglement of two spins by a particular noisy environment. Different types
of noisy environments can be distinguished using this quantum analysis. This approach will allow one to develop new quantum technology to characterize the noisy environments created in biological systems, by measuring the level of entanglement (concurrence) the environments generate between two nuclear spins.

Suppose that an ensemble of two-spin molecules is diluted in a solution filled with microorganisms (e.g. bacteria or viruses). We single out two nuclear spins, which can be studied by conventional NMR techniques at room temperature. So, in our ensemble each two-spin molecule is attached to a single bacterium (virus). We assume that these chosen molecules will attach to bacteria so that the two spins of the attached molecule interact through the noisy environments (local and collective) produced by the bacteria.  One considers three such samples of diluted two-spin molecules attached to bacteria. Three samples are necessary for implementing a spatial labeling \cite{chuang}. (We choose spatial labeling for definiteness, but temporal labeling with a single sample is possible as well.) External electromagnetic pulses can selectively excite each sample. The signals from the three samples are added to obtain their average. In the method of spatial labeling, the spin systems in all samples are initially in equilibrium at room temperature. Then a preliminary sequence of radio-frequency ({\it rf} ) pulses is applied to two of the three samples in order to change their mixed state by permuting diagonal density matrix elements. Then the evolution of the three-sample average NMR signal is exactly the same as if the two spins were initially in their ground states. One can reasonably doubt the possibility of entangling spins which begin their evolution from a state of thermal equilibrium with the noisy environment. However,  labeling (averaging) reveals the evolution of the NMR signal, which corresponds to the initial ground states of the spins. One can consider this spatial labeling as the simulation
of the NMR signals, that would be obtained if the spins were initially in their ground state. Thus, in this case one may consider an ensemble of two-spin systems, which start from the ground state.

After the initial preparation of the samples (permutation of the
density matrix elements), all three samples are subjected to the same sequences of the
{\it rf} pulses. First, one applies initial pulses which change the mutual orientation of the spins. Then, the noisy environment produced by bacteria can generate entanglement. The level of entanglement, which may be negligible for one kind of bacteria and significant for another kind, is determined by measuring the concurrence. To find the latter, one applies, to all three samples, electromagnetic pulses to carry out a quantum state tomography protocol \cite{chuang}. The state tomography protocol allows one to recover the spin density matrix. (More precisely, in order to implement the state tomography
protocol, one repeats the whole procedure nine times with different sequences of state tomography pulses.) Once the spin density matrix is recovered, one is able to compute the concurrence by measuring the entanglement created by the
noisy environment. This determines whether the noise produced
by bacteria creates quantum entanglement in a two-spin system. We assume
that the scalar J-J interaction between the two nuclear spins mediated by the chemical bonds in a molecule does not produce entanglement. To achieve the latter case, the mutual orientation of the spins created by the initial pulses should be chosen in such a way that the J-J interaction does not produce entanglement.

In order to be feasible this approach requires the following: (1) The interaction with the noisy environment must produce significant entanglement during a time shorter than the relaxation and decoherence times. (2) The correlation length
of the noisy environment must be greater than the distance between the two spins in the molecule. (3) The relaxation and decoherence times in the two-spin system must exceed the time required to apply the pulse sequence in the quantum tomography measurement.

Condition (1) imposes a requirement on the effective constant of the spin-environment interaction. Condition (2) imposes a requirement on the correlation radius
of the spin-environment interaction. Condition (3) is satisfied in liquid NMR \cite{chuang}.

As a concrete example one can consider molecules of chloroform \cite{chuang}. In chlorophorm, the frequencies of the hydrogen and carbon nuclear spins are well separated and can be addressed independently. The dipole-dipole interaction between the spins in liquids is suppressed due to the motion of the molecules. The J-J scalar coupling is mediated by chemical bonds with a corresponding frequency of about 200 Hz. In this example, an initial non-equilibrium state should be chosen so that the J-J coupling does not create
its own entanglement with  a significant value of concurrence. The relaxation and dephasing times for the proton spin are 18s and 7s, and for the carbon spin 25s and 0.3s. These times are long enough to implement the two-spin tomography.

We note that any quantum information processor could be used for the study of entanglement induced by a noisy environment. One could consider, for example, two ions in an ion trap quantum computer \cite{chuang} or two superconducting qubits \cite{Martinis}. Our example with the NMR quantum information processor is associated with the possibility to study entanglement induced by biological objects and has as general goals the creation of the novel devices exploiting quantum effects at room temperature. The NMR quantum information processor is, probably, the only recently available technique of this kind.

\medskip

In the present paper, we consider a model of $N$ not directly interacting spins $1/2$ (qubits) placed in a constant effective magnetic field (oriented in the $z$-direction). The qubits interact with both local and collective thermal environments (all at the same temperature). The collective interaction introduces an indirect qubit interaction. 
In the total density matrix of all qubits and environments, we trace over the variables of the environments and $N-2$ qubits. This gives us the time-dependent reduced density matrix for two arbitrary qubits. In the $z$-representation it is represented by a time-dependent $4\times 4$ matrix. It is important to notice that the matrix elements, $[\rho_t]_{n,m}$, $n,m=1,...,4$, depend not only on the parameters of the thermal environments but also on the total number of qubits, $N$. We study numerically the concurrence $C(t)$ of the reduced two-qubit density matrix, and its dependence on the parameters of the system and on $N$. To realize and study this situation in an experiment, one must have access to the two selected qubits (such as their particular frequencies), in order to manipulate them and prepare the initial state. Our main result is that the amplitude of concurrence, $C_{max}(t)$, generally decreases as $N$ increases. This means that one should not expect that the collective thermal environment can create by itself measurable entanglement even of two-qubits, in the presence of many other qubits within a range of the collective environment coherence length.

\subsection{Outline of main results}

The initial state of the entire system is disentangled, a product state in which each of the $N$ spins is in a state $\rho_j$, $j=1,\ldots,N$, all local reservoir states are thermal and so is that of the collective reservoir (at a fixed common temperature \footnote{A generalization to a non-equilibrium situation where each local and the collective reservoir have different individual temperatures is immediate.}).

\medskip
\noindent
{\bf Analytic results.\ }

$\bullet$ {\it Explicit dynamics.\ } As the spins interact with the reservoirs via energy-conserving couplings only, the reduced two-spin dynamics can be calculated explicitly, see Proposition \ref{prop1} below. Consequences of the energy conservation are that populations, i.e., the diagonal density matrix elements, are time-independent, and that the off-diagonal elements evolve independently. As an example, we discuss here the dynamics of the (1,2) matrix element,
\begin{equation}
{}[\rho_t]_{12} = [\rho_0]_{12}\ \ \e^{\i\omega_2t}\ \e^{\i\varkappa^2_c S(t)}\ \e^{-\varkappa^2_\ell\Gamma_\ell(t) -\varkappa^2_c\Gamma_c(t)}\  P_N(t).\label{1.14'}
\end{equation}
The other matrix elements have similar behavior. Each factor on the r.h.s. has an interpretation:

-- $[\rho_0]_{12}$ is the initial condition of the matrix element in question. None of the other initial matrix elements are involved (energy conserving coupling);

-- $\e^{\i\omega_2t}$ is the uncoupled dynamics (no interaction with environments);

-- $\e^{\i\varkappa^2_c S(t)}$ is a dephasing factor with a time-dependent phase $S(t)\leq 0$ becoming linear for large $t$ (for the considered infra-red behavior $|k|^{1/2}$ of the coupling constants in three dimensions, see \fer{soft}); it represents a ``Lamb shift'' contribution to the real part of the effective energy; this term is generated by the collective reservoir, but it is independent of the presence of the $N-2$ traced-out spins (the term would be the same if only two spins were coupled to the reservoirs);

-- $\e^{-\varkappa^2_\ell\Gamma_\ell(t) -\varkappa^2_c\Gamma_c(t)}$ is a decaying factor with time-dependent decay rates, $\Gamma(t)\geq 0$, becoming linear for large $t$ (see \fer{gammaoft}) both the local and collective reservoirs contribute; however, the term is independent of the $N-2$ traced-out spins (again, it would be the same if only two spins were coupled to the reservoirs);

-- $P_N(t)$ is a product of $N-2$ oscillating terms encoding the effect of all the traced-out spins (see \fer{1.11}). It is important to notice that $P_N(t)$ {\it only depends on the diagonal density matrix elements of the initial states of the $N-2$ traced-out spins}.\footnote{This is so since the dynamics is energy-conserving, and tracing out any of the spins involves only the diagonal of the initial (time zero) density matrix. (See also Remark 1 after Proposition \ref{prop1}.)} Consequently, the two-qubit state does not depend on the initial off-diagonal density matrix elements of the $N-2$ traced-out ``background'' spins. Typically, we expect those spins to be initially in (close to) equilibrium, corresponding to vanishing off-diagonals.

Some general properties of $P_N(t)$ can be explained easily for the case in which all $N-2$ spins are initially in the high temperature equilibrium state $\frac{1}{\sqrt{2}} ( |+\rangle +|-\rangle)$. Then $P_N(t)=[\cos(\varkappa_c^2S(t))]^{N-2}$ and its magnitude oscillates between zero and one. Due to the large power $(N-2)$, the peaks of the function $|P_N(t)|$, centered at the discrete times $t_p$ satisfying $S(t_p)\in \frac{\pi}{\varkappa^2_c}\mathbb Z$, are of very narrow width $O(1/(\varkappa_c^2\sqrt{N}))$ for large $N$. Consequently, in the limit $N\rightarrow\infty$, with $\varkappa_c$ held fixed, $|P_N(t)|$ is zero for all $t$ except for $t=t_p$, where $|P_N(t_p)|=1$. But the density matrix becomes very simple if $P_N(t)=0$, because many entries vanish (c.f. Proposition \fer{prop1}) and the corresponding concurrence is zero. It follows that in the large $N$ limit, concurrence is zero for all times (except possibly for some isolated instances, $t_p$).

$\bullet$ {\it $N$-dependent scaling of the interaction $\varkappa_c$.\ } The above analysis suggests that one cannot generate two-spin entanglement for $N$ large at fixed interaction strength $\varkappa_c$. However, the width of the peaked function $P_N(t)$ which is of order $O(1/(\varkappa^2_c\sqrt{N}))$ becomes appreciable if $\varkappa_c\gtrsim N^{1/4}$. Hence we consider a $N$-dependent scaling of the coupling, replacing $\varkappa_c$ by $\varkappa_c/N^\eta$, for some $\eta>0$. According to the above discussion, the borderline case is $\eta=1/4$.

Starting from the explicit expressions (Proposition \ref{prop1}) and using the scaling $\varkappa_c/N^\eta$, we calculate the limit $N\rightarrow\infty$  of $\rho_t$, for $t\in\rx$ fixed. The analytic expressions we obtain for $0<\eta<1/4$ and $\eta>1/4$ show that the limiting dynamics does not create entanglement, for any time $t$. While we are able to obtain explicit expressions for concurrence in the regime of $N\rightarrow\infty$, we are not so for $N$ finite.\footnote{The reduced density matrix is given explicitly for all $N$ and all $t$, but calculating from it explicitly the concurrence is more difficult.} However, since no entanglement is generated in the limiting case, $N\rightarrow\infty$, but we know entanglement is created for $N=2$ (see e.g. \cite{Braun,MBBG}), we expect that entanglement creation decays with increasing $N$. We study this decay numerically.

\medskip
\noindent
{\bf Numerical results.\ }

We introduce $\nu_c$, the highest frequency at which spin-reservoir interactions occur and
call it the cutoff frequency. In the simulations, we take $\nu_c$ of the order of
the thermal frequency $\nu_T=k_BT/h$. In the infra-red regime, our coupling is
 proportional to $\sqrt{|k|}$ (see after \ref{gammaoft}).

$\bullet$ For $N=2$, concurrence creation is maximal if both spins
start out in their high-temperature state $\frac{1}{\sqrt{2}}(|+\rangle+|-\rangle)$,
see Fig.1. Consequently, in the subsequent simulations, we take initial states of the two not traced-out qubits very
close to this state, and we take the diagonals of the initial states of the $N-2$ traced-out quibts to be constant $1/2$ (remember that the off-diagonals of these qubits do not influence the dynamics at all). In Fig.4 we modify the initial state of the two not traced-out qubits and check that maximal concurrence is indeed obtained when both qubits are in the above state, even for large $N$.

$\bullet$ For general $N$, entanglement evolves according to a rescaled
time $t\mapsto \varkappa^2_c\nu_ct$, see Fig.2. This figure shows that a
reduction of $\varkappa_c$ diminishes the created concurrence in a moderate way.
For instance decreasing $\varkappa_c$ by a factor $10$ only decreases concurrence by less than $1/3$.

$\bullet$ In Fig.3 we show that the maximum of created concurrence decays with increasing $N$.
For intermediate values of $N$ (with the current parameters $N\sim 10-150$) the decrease
 is exponential, for smaller and larger values of $N$, it is superexponential.

$\bullet$ In the same Fig.3c, we study the dependence of the maximal time, $\tau_c$,  (before recurrence) at
which the concurrence is not zero. We have found that this time decays exponentially in the number of spins, $N$,
for sufficiently large $N$.

$\bullet$ Results on the rescaled model $\varkappa_c\mapsto \varkappa_c/N^\eta$ are shown
in Fig.5. We find a decrease of maximal concurrence with increasing $N$ for
all $\eta\geq 0$. The critical value, $\eta=1/4$ (see analytic results above)
divides the concurrence decay into two regimes. In the range, $\eta>1/4$, the maximal
concurrence decreases exponentially in $N$, for intermediate values of $N$ (between 10 and 180),
with a {\it universal decay rate}
(i.e., not depending on $\eta$). For $\eta<1/4$ the decay is superexponential
and varies with $\eta$. We conclude that no scaling $\varkappa_c\mapsto \varkappa_c/N^\eta$ can
compensate the decay of created concurrence for large $N$.

\section{Model and reduced density matrix}

The full Hamiltonian of the $N$ noninteracting spins $1/2$ coupled by energy conserving interactions to local and collective bosonic heat reservoirs is given by
\begin{eqnarray}
H &=& -\hbar \sum_{n=1}^N\omega_n S_n^z +\sum_{n=1}^N H_{\r_n} +H_\r \label{1}\\
 &&+\sum_{n=1}^N\varkappa_n S_n^z\otimes\phi_\c(f_\c)+ \sum_{n=1}^N\nu_n S_n^z\otimes\phi_n(f_n). \label{2}
\end{eqnarray}

Below we use dimensionless variables and parameters. To do so, we
introduce a characteristic frequency, $\omega_0$, typically of the
order of spin transition frequency. The total Hamiltonian, energies
of spin states, and temperature are measured in units
$\hbar\omega_0$. The frequencies of spins, $\omega_n>0$, bosonic
excitations, $\omega(k)=c|\vec{k}|$ (where $c$ is the speed of
light), the wave vectors of bosonic excitations are normalized by
$\omega_0/c$, and all constants of interactions are measured in
units of $\omega_0$.  The dimensionless time is defined as $\omega_0t$.

In \fer{1}, \fer{2}, $\omega_n>0$ is the frequency of spin $n$,
\begin{equation}
S^z = \frac{1}{2}
\left[
\begin{array}{cc}
1 & 0\\
0 & -1
\end{array}
\right]
\label{sz},
\end{equation}
and $S^z_n$ denotes the $S^z$ of spin $n$. $H_\r$ is the
Hamiltonian of the bosonic collective reservoir,
\begin{equation}
H_\r = \int_{\rx^3} |k| a^*(k)a(k) \d^3k, \label{n2}
\end{equation}
and $H_{\r_n}$ is that same Hamiltonian for the $n$-th
individual reservoir.  For a square-integrable {\it form factor}
$h(k)$, $k\in\rx^3$,  $\phi(h)$ is given by
\begin{equation}
\phi(h) = \frac{1}{\sqrt 2}\int_{\rx^3}\left\{ h(k) a^*(k) + h(k)^* a(k)\right\} \d^3k.
\label{phi}
\end{equation}
The real numbers, $\varkappa_n$ and $\nu_n$,  are
coupling constants, measuring the strengths of the energy-conserving collective
coupling and the energy-conserving local coupling, respectively.



Since the spins interact with the reservoirs only through energy-conserving channels,
this model is {\it exactly solvable}. For simplicity of exposition, we take
$$
\varkappa_n = \varkappa_c \mbox{\quad for all $n$ (collective)}
$$
and
$$
\nu_n = \nu_\ell \mbox{\quad for all $n$ (local)}.
$$
We also take, for simplicity, all local form factors equal ($f_\ell$) and all collective ones also ($f_c$).

Fix any pair of spins, and (re-)label their frequencies by $\omega_1$ and $\omega_2$, see \fer{1}. We write the reduced density matrix, $\rho_t$, of the two fixed spins as a $4\times 4$ matrix $[\rho_t]_{ij}$ in the ordered energy basis
\begin{equation}
\Phi_1=\varphi_1\otimes\varphi_1,\ \  \Phi_2=\varphi_1\otimes\varphi_{-1},\ \  \Phi_3=\varphi_{-1}\otimes\varphi_1, \ \ \Phi_4=\varphi_{-1}\otimes\varphi_{-1},\footnote{Another equivalent notation is: $\Phi_1=|++>$, $\Phi_2=|+->$, $\Phi_3=|-+>$, $\Phi_4=|-->$.}
\label{basis}
\end{equation}
where $S^z\varphi_{\pm 1}=\pm\frac12\varphi_{\pm 1}$. For instance, $[\rho_t]_{2,4} = \scalprod{\Phi_2}{\rho_t\Phi_4}$.

The {\it initial state} of the spins is a product state
of the form $\rho_{S_1,0}\otimes\cdots\otimes \rho_{S_N,0}$, where
\begin{equation}
\rho_{S_j,0} =
\left[
\begin{array}{cc}
p_j & v_j\\
v_j^* & 1-p_j
\end{array}
\right],
\label{initstate}
\end{equation}
with $0\leq p_j\leq 1$ and $|v_j|^2\leq p_j(1-p_j)$. The upper bound on the off-diagonal guarantees that the eigenvalues of $\rho_{S_j,0}$ are non-negative.

We introduce the quantities
\begin{eqnarray}
P_N(t) &=& \prod_{j=3}^N \left[ p_j\ \e^{\i\varkappa^2_c S(t)} +(1-p_j)\
 \e^{-\i\varkappa^2_c S(t)}\right],\label{1.11}\\
S(t) &=& -\frac12 \int_{\rx^3}|f_c(k)|^2 \frac{|k|t -\sin(|k|t)}{|k|^2} \d^3k,
\label{soft}\\
\Gamma_{\ell,c}(t) &=&\int_{\rx^3} |f_{\ell,c}(k)|^2
\coth(\beta|k|/2) \frac{\sin^2(|k|t/2)}{|k|^2} \d^3k.
\label{gammaoft}
\end{eqnarray}

The integrals in \fer{soft} and \fer{gammaoft} are made to converge introducing a suitable cut-off wavenumber, $|k_c|$, or cut-off frequency, $\nu_c = |k_c|/2\pi$.
(Here we use dimensionless units.) For instance, for numerical simulations, we choose as form factor the function $f_c (k) = \sqrt{|k|}\ \chi_{|k|\leq |k_c|}$, where $\chi_{|k|\leq |k_c|}=1$ if $|k|\leq |k_c|$ and $\chi_{|k|\leq |k_c|}=0$ otherwise.

We also define $\widetilde P_N(t)$ to be the same as $P_N(t)$, but with $\varkappa^2_c$ replaced by $2\varkappa^2_c$. With this notation, we have the following result.
\begin{prop}[Explicit dynamics of the reduced density matrix.]
\label{prop1}
The evolution of the density matrix is given  by
\begin{eqnarray}
{}[\rho_t]_{12} &=& [\rho_0]_{12}\ \ \e^{\i\omega_2t}\ \e^{\i\varkappa^2_c S(t)}\ \e^{-\varkappa^2_\ell\Gamma_\ell(t) -\varkappa^2_c\Gamma_c(t)}\  P_N(t),\label{1.14}\\
{}[\rho_t]_{13} &=& [\rho_0]_{13}\ \  \e^{\i\omega_1t}\ \e^{\i\varkappa^2_c S(t)}\  \e^{-\varkappa^2_c\Gamma_c(t)}\  P_N(t),\\
{}[\rho_t]_{14} &=& [\rho_0]_{14}\ \ \e^{\i(\omega_1+\omega_2)t}\ \ \e^{-2\varkappa^2_\ell\Gamma_\ell(t) -4\varkappa^2_c\Gamma_c(t)}\ \widetilde P_N(t),\\
{}[\rho_t]_{23} &=& [\rho_0]_{23}\ \ \e^{\i(\omega_1-\omega_2)t}\ \e^{-2\varkappa^2_\ell\Gamma_\ell(t)},\\
{}[\rho_t]_{24} &=& [\rho_0]_{24}\ \ \e^{\i\omega_1t}\ \e^{-\i\varkappa^2_c S(t)}\ \e^{-\varkappa^2_\ell\Gamma_\ell(t) -\varkappa^2_c\Gamma_c(t)}\  P_N(t),\\
{}[\rho_t]_{34} &=& [\rho_0]_{34}\ \ \e^{\i\omega_2t}\ \e^{-\i\varkappa^2_c S(t)}\ \e^{-\varkappa^2_\ell\Gamma_\ell(t) -\varkappa^2_c\Gamma_c(t)}\  P_N(t),
\label{1.19}
\end{eqnarray}
and the populations are constant, $[\rho_t]_{jj} = [\rho_0]_{jj}$, for $j=1,\ldots,4$ and $t\in\rx$.
\end{prop}

 The proof of this proposition is a rather simple calculation. One can proceed as in \cite{MSB1} (proof of Proposition 7.4).

\medskip

{\bf Remarks.\ } {\bf 1.} The effect of spins $3,\ldots,N$ is contained entirely in the factors $P_N(t)$ and
$\widetilde P_N(t)$. They only depend on the initial populations $p_j$, $j=3,\ldots,N$ (see \fer{1.11}), but not on the off-diagonals, $v_j$. This is explained by the fact that when tracing over a single spin, $j\geq 3$, we perform the operation ${\rm Tr}_{{\rm spin} j}\ U \rho_{S_j,0}V$, where $U,V$ are operators commuting with $S_2^z$ (energy conserving interactions only!). Clearly the latter trace only involves the diagonal of $\rho_{S_j,0}$.

{\bf 2.} The oscillatory phases, $\e^{\i \omega t}$, in \fer{1.14}-\fer{1.19} represent the free, uncoupled dynamics of the spins. Consider the modified two-spin density matrix
\begin{equation}
\rho'_t=\e^{\i t(-\omega_1 S^z_1 -\omega_2 S^z_2)}\ \rho_t\  \e^{-\i t(-\omega_1 S^z_1 -\omega_2 S^z_2)},
\label{rhoprime}
\end{equation}
(``interaction picture'' dynamics of $\rho_t$). Because $\rho'_t$ and $\rho_t$ are related by conjugation of a unitary operator of the product form, $\e^{\i t(-\omega_1 S^z_1)}\otimes\e^{\i t(-\omega_2 S^z_2)}$, {\it the concurrences of $\rho_t$ and $\rho'_t$ are the same}. In other words, when examining concurrence of $\rho_t$, we may use formulas \fer{1.14}-\fer{1.19} with $\omega_1=\omega_2=0$.

\subsection{Concurrence}

Recall that the concurrence of the reduced density matrix, $\rho_t$, is unchanged when we pass to the interaction picture $\rho'_t$ (see the remark explaining \fer{rhoprime}).
In the basis \fer{basis}, the evolution of $\rho'_t$, \fer{rhoprime}, is given by \fer{1.14}-\fer{1.19} with $\omega_1=\omega_2=0$, and where the initial condition is ($\rho'_0=\rho_0$)
\begin{equation}
[\rho_0]=
\left[
\begin{array}{cccc}
p_1p_2  &  p_1v_2  &  v_1p_2  &  v_1v_2\\
p_1v^*_2  &  p_1(1-p_2)  &  v_1v^*_2  &  v_1(1-p_2)\\
v^*_1p_2  &  v^*_1v_2  &  (1-p_1)p_2  &  (1-p_1)v_2\\
v^*_1v^*_2  &  v^*_1 (1-p_2)  &  (1-p_1)v^*_2  &(1-p_1)(1-p_2)
\end{array}
\right].
\end{equation}

\subsubsection{Variation of $N$-dependence}

For homogeneous initial conditions, $p_j=p$ for $j=3,\ldots,N$, we have
\begin{equation}
P_N(t) = [p\e^{\i\varkappa_c^2 S(t)}+(1-p)\e^{-\i\varkappa^2_c S(t)}]^{N-2}.
\label{1.22}
\end{equation}
Unless $p=0,1$, $|P_N(t)|$ oscillates in $t$ between its minimum value $|2p-1|^{N-2}$ (when $\cos(2\varkappa_c^2S(t))=-1$) and its maximum value $1$ (when $\cos(2\varkappa_c^2S(t))=1$).  \footnote{We have $|p\e^{\i\varkappa_c^2 S(t)}+(1-p)\e^{-\i\varkappa^2_c S(t)}|^2=p^2+2p(1-p)\cos(2\varkappa^2_cS(t)) +(1-p)^2$.} The width of the oscillations becomes very narrow with increasing $N$. In the limit of large $N$, $P_N(t)$ is zero for all times, except for the discrete set of $t\in\rx$ satisfying $\cos(2\varkappa_c^2S(t))=1$, in which case $|P_N(t)|=1$.

This implies that for large $N$, all off-diagonal density matrix elements of $\rho_t$ vanish with the exception of $[\rho_t]_{23} = \e^{\i(\omega_1-\omega_2)t}\e^{-2\varkappa^2_\ell\Gamma_\ell(t)}$ (and $[\rho_t]_{32}$, of course) for almost all values of $t$. This suppression of off-diagonals comes from the large number of particles and is mediated through the collective energy-conserving interaction. (For $\varkappa_c=0$ we have $P_N(t)=1$.)

In order to try to have a non-trivial dynamics for large $N$, one may scale the collective conserving coupling constant as
$$
\varkappa_c\rightarrow \frac{\varkappa_c}{N^\eta},\qquad \mbox{some $\eta >  0$}.
$$
Then \fer{1.22} becomes
\begin{equation}
P_N(t) = [p\e^{\i\varkappa_c^2 S(t)/N^{2\eta}}+(1-p)\e^{-\i\varkappa^2_c S(t)/N^{2\eta}}]^{N-2}.
\label{1.23}
\end{equation}
An expansion in large $N$ yields
$$
P_N(t) =\e^{-\i\varkappa^2_cS(t)[1-2p] N^{1-2\eta}}\ \e^{-2\varkappa^4_cS^2(t) N^{1-4\eta}[p(1-p) + O(N^{-2\eta})]}.
$$
Thus as $N\rightarrow\infty$,
\begin{equation}
P_N(t)\rightarrow \left\{
\begin{array}{ll}
\e^{-\i\varkappa^2_cS(t)[1-2p] N^{1/2}}\ \e^{-2\varkappa^4_cS^2(t)p(1-p)N^{1-4\eta}} & 0<\eta\leq 1/4  \mbox{\quad (and $p\neq 0,1$})\\
\e^{-\i\varkappa^2_cS(t)[1-2p]N^{1-2\eta}} & 1/4<\eta
\end{array}
\right.
\label{m1}
\end{equation}
{\bf Remarks.\ } {\bf 1.\ } By replacing, in these limits, $\varkappa^2_c$ by $2\varkappa^2_c$, we obtain the corresponding limits for $\widetilde P_N(t)$.

{\bf 2.\ } For $0<\eta<1/4$, $P_N(t)$ vanishes as $N\rightarrow\infty$.

{\bf 3.\ } The rapid oscillating phases disappear if $p=1/2$ (any $\eta>0$) or $\eta=1/2$ (any $p$).

\subsubsection{Asymptotic concurrence  ($N\rightarrow\infty$)}

$\bullet$  $0<\eta<1/4:$ \quad The reduced two-spin density matrix (in the interaction picture) at time $t$  is
\begin{equation}
[\rho'_t]=
\left[
\begin{array}{cccc}
p_1p_2  &  0  & 0 &  0\\
0 &  p_1(1-p_2)  &  v_1v^*_2\e^{-2\varkappa_\ell^2\Gamma_\ell(t)}  &  0\\
0 &  v^*_1v_2\e^{-2\varkappa_\ell^2\Gamma_\ell(t)}  &  (1-p_1)p_2  &  0\\
0  &  0 &  0 &(1-p_1)(1-p_2)
\end{array}
\right],
\label{1.25'}
\end{equation}
from which we obtain the concurrence
\begin{equation}
C(\rho_t) = \max\left\{0, -2[\sqrt{p_1(1-p_1)p_2(1-p_2)} -|v_1|\, |v_2| \e^{-2\varkappa^2_\ell\Gamma_\ell(t)}] \right\} = 0.
\end{equation}
Remember that $|v_j|^2\leq p_j(1-p_j)$. This shows that the $N\rightarrow\infty$ asymptotic dynamics cannnot create entanglement at any time.



\medskip

$\bullet$ $\eta>1/4:$ \quad Call the r.h.s. of \fer{m1} $P_\infty(t)$ (a quantity still depending on $N$ unless $\eta=1/2$). By replacing $\varkappa^2_c$ by $2\varkappa^2_c$ in \fer{m1} we obtain the limit of $\widetilde P_N(t)$, which we call $\widetilde P_\infty(t)$. For  $\eta>1/4$ we have the relation $\widetilde P_\infty(t) = [P_\infty(t)]^2$.  The reduced two-spin density matrix (in the interaction picture) at time $t$ is, for $N\rightarrow\infty$
\begin{eqnarray}
\lefteqn{
[\rho'_t]=}\label{1.25} \\
&&
\left[
\begin{array}{cccc}
p_1p_2  &  p_1v_2D_\ell P_\infty(t)  &  v_1p_2 D_\ell P_\infty(t) &  v_1v_2D_\ell^2P_\infty(t)^2\\
p_1v^*_2D_\ell P_\infty(t)^*  &  p_1(1-p_2)  &  v_1v^*_2D_\ell^2  &  v_1(1-p_2)D_\ell P_\infty(t)\\
v^*_1p_2 D_\ell P_\infty(t)^* &  v^*_1v_2 D_\ell^2 &  (1-p_1)p_2  &  (1-p_1)v_2D_\ell P_\infty(t)\\
v^*_1v^*_2D_\ell^2[P_\infty(t)^2]^*  &  v^*_1 (1-p_2)D_\ell P_\infty(t)^*  &  (1-p_1)v^*_2D_\ell P_\infty(t)^*  &(1-p_1)(1-p_2)
\end{array}
\right],
\nonumber
\end{eqnarray}
where
$$
D_\ell =D_\ell(t)= \e^{-\varkappa^2_\ell \Gamma_\ell(t)},\qquad P_\infty(t) = \e^{-\i\varkappa_c^2 S(t)[1-2p]N^{1-2\eta}}.
$$
The density matrix \fer{1.25} is of the product form
$$
\left[
\begin{array}{cc}
p_1 & v_1D_\ell(t) P_\infty(t)\\
v_1^*D_\ell(t) P_\infty(t)^* & 1-p_1
\end{array}
\right]
\otimes
\left[
\begin{array}{cc}
p_2 & v_2D_\ell(t) P_\infty(t)\\
v_2^* D_\ell(t) P_\infty(t)^*& 1-p_2
\end{array}
\right].
$$
This shows that the $N\rightarrow\infty$ asymptotic dynamics is factorizable and cannnot create entanglement at any time.

\section{Numerical Results}

{\ }
\indent
$\bullet$ Let us first consider the case of two spins only, $N=2$.
 In \fer{1.14}-\fer{1.19} we put for simplicity
 $\Gamma_\ell = \Gamma_c = \Gamma $ and regard  $\Gamma$  and $S$ as
two independent parameters. Taking both spins initially in the same
state given by $p$, $v$, see \fer{initstate}, we examine the maximal
concurrence, as a function of $S$ and $\Gamma$, for arbitrary
fixed values of $p$ and $v$.

\begin{figure}[!ht]
\begin{center}
\epsfxsize=55mm
\epsfbox{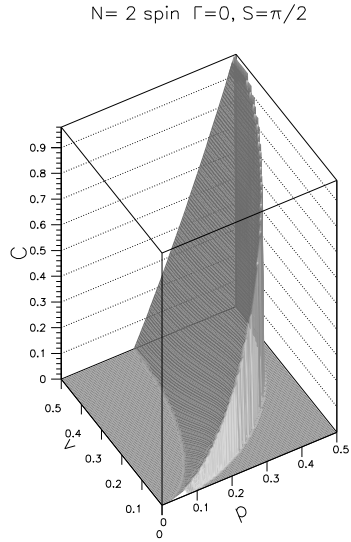}
\end{center}
\caption{{\small
Maximal concurrence as a function of $p$ and $v$, for fixed
$\Gamma(t)=0$ ans $S(t)=\pi/2$. Here, $N=2$ spins is considered.
}}
\label{uno}
\end{figure}

We find that  for fixed $p$, $v$, the maximal concurrence is
given at $S=\pi/2,~ \Gamma=0$. Having such values fixed and
plotting the concurrence as a function of $p,v$, the maximal
concurrence is realized when $p=v=1/2$, see Fig.~\ref{uno}, where a plot of
the concurrence as a function of $p,v$
is shown. Maximal generation of concurrence is thus obtained starting
from pure state initial conditions $\frac{1}{\sqrt{2}}(|+\rangle+|-\rangle)$
for each spin.

$\bullet$ Let us now  consider the case of $N$ spins. For concreteness we choose $p=1/2$ for all spins (the traced-out ones and the two not traced-out ones). For the two not traced-out spins we take off-diagonals $v=0.48$. (Then $p$ is close to $v$ which favors larger entanglement creation). Recall that the dynamics is independent of the off-diagonals of the $N-2$ traced-out spins (i.e., we do not have to specify the $v$ of the $N-2$ traced-out spins).

As mentioned after \fer{gammaoft}, we choose the form factor,
$f_c (k) = \sqrt{|k|}\ \chi_{|k|\leq2\pi \nu_c}$, with the cut-off frequency equal
to the thermal frequency, $\nu_c= \nu_T=  k_B T/h$, at room temperature, $T=300$ K.

\begin{figure}[ht]
\begin{center}
\epsfxsize=75mm
\epsfbox{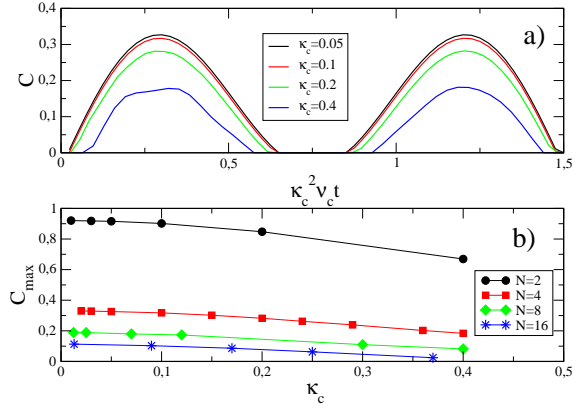}
\end{center}
\caption{{\small
a) Concurrence as a function of the rescaled time, for
different $\varkappa_c$ and  fixed
$\epsilon = \nu_c/\nu_T = 1$, being respectively the cut-off and thermal
frequency ($\nu_T = k_B T /h$ with  $T=300 K$) and the form factor
$f_c(k ) = \sqrt{|k|}$. Other data are $p=0.5$, $v=0.48$, $N=4$.
b) Plot of the concurrence at the peak (obtained from a)) as a function
of the coupling strength, $\varkappa_c$, for different $N$ values as indicated
in the legend. Other values are the same as in a) .
}}
\label{prima}
\end{figure}

In Fig.~\ref{prima} we investigate the effects of an increase in
the coupling parameter, $\varkappa_c$. The first effect
is a time-shift for the concurrence evolution, described
by a scaling, $t \to \varkappa_c^2 \nu_c t$,
 see Fig.~\ref{prima}a,
where $\nu_c$ is the cut-off frequency.
The second effect is a reduction
of the maximal concurrence in a smooth way,
see Fig.~\ref{prima}b.
As one can see, the effective decrease in amplitude for not
too strong coupling strength, $\varkappa_c$, is relatively small.
For instance, changing $\varkappa_c$ for $N=2$ by one order of magnitude
from 0.04 to 0.4, changes the amplitude by only 27\%.
The percentage change is almost the same for larger $N$ values,
see Fig.~\ref{prima}b.

\begin{figure}[ht]
\begin{center}
\epsfxsize=85mm
\epsfbox{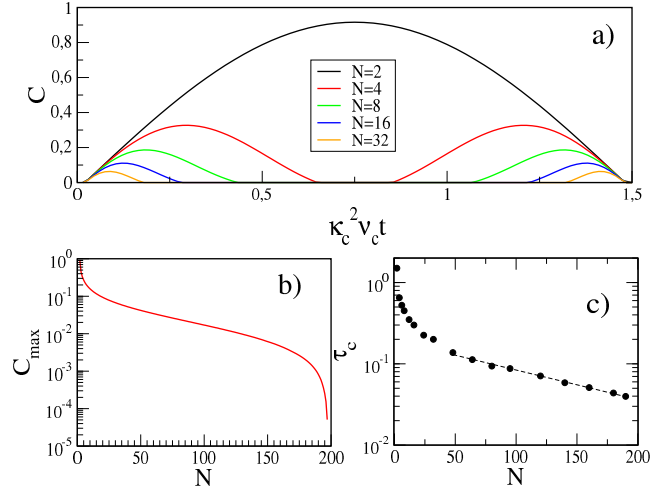}
\end{center}
\caption{{\small
a) Concurrence as a function of the rescaled time, for fixed
$\epsilon = \nu_c/\nu_T = 1$, being respectively the cut-off and thermal
frequency ($\nu_T = k_B T /h$ with  $T=300 K$) and the form factor,
$f_c(k ) = \sqrt{|k|}$. Other data are $p=0.5$, $v=0.48$, $\varkappa_c=0.05$.
b) Plot of the concurrence at the peak (obtained from a)) as a function
of the number of spins $N$.
c) Plot of the collapse time, $\tau_c$, as a function of the number of spins, $N$.
The dashed line is the best fit exponential $\exp(-\alpha N)$ with $\alpha = 0.0838\pm 0.0002$.
}}
\label{quattro}
\end{figure}

In Fig. \ref{quattro} we show that the creation of concurrence
decreases with the number of spins. In graph a) we plot the
concurrence as a function of (rescaled) time for various values of
$N=2,\ldots,32$.  As one can see,
the same time rescaling is also valid for $N>2$.
Moreover, the maximum concurrence created, $C_{\rm max}$, reported
in graph b), decreases exponentially in $N$ in the range  $10 < N < 150 $
and faster than exponentially outside this range.
 For larger $N$, the  concurrence
decays superexponentially in $N$. For $N$ exceeding $200$, the concurrence becomes too small to be
 significant (of the order $10^{-4}$).

It is also interesting to note that the graph of concurrence shows collapses and revivals and that
the revival times for  $N>2$ are always less than the revival time for $N=2$. It is also interesting to consider how the collapse time, $\tau_c$, defined as the first time at which
concurrence drops abruptly to zero, depends on the number of spins, $N$. This study has been reported in Fig.~\ref{quattro}c and shows that the rescaled collapse time
decays exponentially with the number of spins: $\tau_c = \varkappa_c^2 \nu_c t_c
\simeq \exp (-\alpha N)$, where $\alpha = 0.0838\pm 0.0002$. (See dashed line in  Fig.~\ref{quattro}c.)

$\bullet$ One can also vary the initial conditions for the spins by choosing independent $p_{1,2}$ and $v_{1,2}$,
while all other spins have the same value $p_j=1/2$, $j=3,\ldots,N$ (their off-diagonals $v_j$ do not influence the dynamics at all).

In order to simplify the problem,  we also set $p_1=v_1$ and
$p_2=v_2$ and consider the maximal concurrence as a function of
two independent parameters, $p_1$ and $p_2$, only.

\begin{figure}[ht]
\begin{center}
\epsfxsize=85mm
\epsfbox{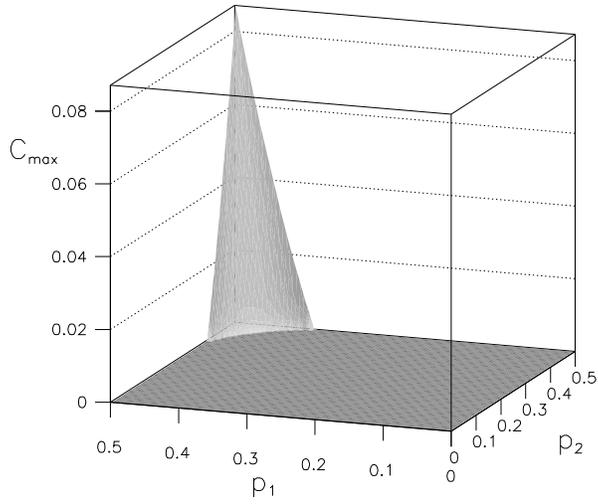}
\end{center}
\caption{{\small
a) Maximal concurrence as a function of the independent
parameters, $p_1=v_1$ and $p_2=v_2$. As one can see the maximal
concurrence is realized at the external corner, i.e $p_1=p_2=1/2$.
Here is $\epsilon=1$, $N=40$ and $p=1/2$ for all other spins.
}}
\label{newentry}
\end{figure}

An example of the 3D plot obtained is reported in Fig.~\ref{newentry} :
 the maximal concurrence is realized for $p_1=p_2=v_1=v_2 = 1/2$, independently of $N$
(in the picture $N=40$, but similar plots are obtained for other values of $N$).

$\bullet$ The numerical analysis of the rescaled model with
$\varkappa_c$ replaced by $\varkappa_c/N^\eta$ shows that the
 concurrence is always a decreasing function of $N$ and that the
 maximum of the created concurrence is a universal function of
the number of spins $N$, independent of $\eta$ for $\eta >1/4$.

\begin{figure}[ht]
\begin{center}
\epsfxsize=85mm
\epsfbox{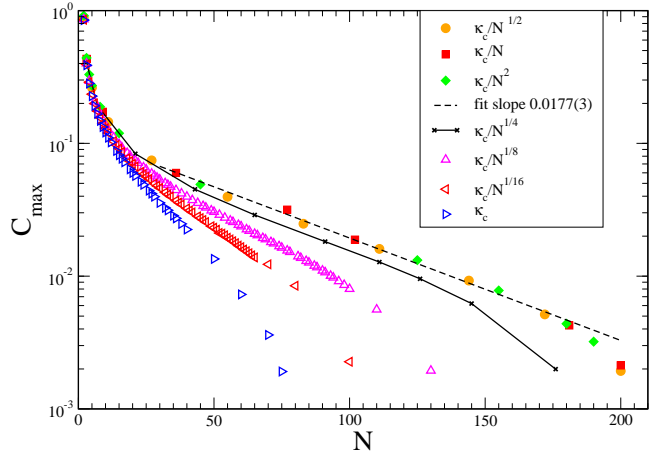}
\end{center}
\caption{
Maximal concurrence as a function of the number of spins, for different
power law scaling, as indicated in the legend. Here is
$f_c(k) = \sqrt{|k|}$, $p=0.5$, $v=0.48$, $\epsilon = \nu_c/\nu_T = 1$,
$T= 300 $ K, $\varkappa_c=0.2$. The dashed line indicates a fitting
exponential for the cases $\eta >1/4$. The solid curve indicates
the case, $\eta =  1/4$.}
\label{cinque}
\end{figure}

 Results are shown in Fig.~\ref{cinque}, where the
dashed line is the best exponential
 fit $\exp( -a N )$, with $a = 0.0177\pm 0.0003$, the best fitting value,
for the cases $\eta >  1/4$. The same figure shows that when $\eta \leq 1/4$
the decay is superexponential and no universality occurs.

This suggests that no power law scaling with $N$ of the coupling strength
can compensate the rapid decay of concurrence with the number of spins.

\noindent
Qualitatively similar results, not reported here,
 can be obtained by changing the ratio between
the thermal and cut-frequency in the range $(0.5,4)$.

\section{Conclusion}

We present a new way to characterize a noise source by analyzing its ability to create
entanglement between two arbitrary qubits in the $N$-qubit open system. We have discussed an application of this approach to the analysis of noise produced by bacteria or viruses. As a first step we consider the dependence of the concurrence on the number of qubits. We show that concurrence quickly decays with increasing number of surrounding qubits. It follows that
for implementing our approach, one has to use a small number of qubits collectively interacting with the thermal environment, preferably only two qubits.


\noindent


\noindent

\noindent


\noindent

\end{document}